\documentstyle[twoside,fleqn,espcrc2,epsf]{article}

\def\ga{\mathrel{\raise.3ex\hbox{$>$\kern-.75em\lower1ex\hbox{$\sim$}}}}
\def\la{\mathrel{\raise.3ex\hbox{$<$\kern-.75em\lower1ex\hbox{$\sim$}}}}
\def\gev{{\rm \, Ge\kern-0.125em V}}
\def\tev{{\rm \, Te\kern-0.125em V}}
\def\msf{m_{\tilde f}}
\def\ohsq{\Omega_{\tilde\chi} h^2}
\def\m12{m_{1\!/2}}
\def\mb{m_{\widetilde B}}
\def\thm{\theta_\mu}
\def\tha{\theta_A}
\def\cp{C\!P}
\def\bino{\widetilde B}
\def\char{\widetilde W}
\def\neut{\tilde\chi^0}
\def\gluino{\tilde g}
\title{$\cp$-Violating Phases in the MSSM}

\author{T. Falk\address{School of Physics and Astronomy, University
                        of Minnesota, Minneapolis, MN 55455, USA}
 \thanks{To appear in the proceedings of SUSY-96: Supersymmetry and 
         Unification of Fundamental Interactions, College Park,
         Maryland, USA 1996}
       }

\begin{document}      

\begin{abstract}
  We combine experimental bounds on the electric dipole moments of the
  neutron and electron with cosmological limits on the relic density
  of a gaugino-type LSP neutralino to constrain certain
  $\cp$-violating phases appearing in the MSSM.  We find that in the
  Constrained MSSM, the phase $|\thm |\la\pi/10$, while the phase
  $\tha$ remains essentially unconstrained.
\end{abstract}

\maketitle


The Minimal Supersymmetric Standard Model (MSSM) contains several new
sources for $\cp$-violation not present in the Standard Model,
and it is well known \cite{dgh,edms,ko} that these phases
can produce large SUSY contributions to the electric dipole moments
(EDM's) of the neutron and electron.  The common generic description
is that these contributions will exceed the current experiment limits
on the neutron and electron EDM's \cite{nexp,eexp} unless either the
$\cp$-violating phases are tiny ($\theta<0.01$) or the sfermion masses
are very large ($\msf>1\tev$).  However, large sfermion masses may be
incompatible with bounds on the relic density $\ohsq$ of a
gaugino-type LSP neutralino.  By combining cosmological and EDM
constraints, we wish to find an upper bound on the magnitude of
$\cp$-violating phases within the MSSM.


In the MSSM, the Higgs mixing mass $\mu$, the guagino mass parameter
$\m12$, the scalar Higgs mixing parameter $B\mu$, and the trilinear
couplings $A$ are all potentially complex.  However, not all of these
phases are physical, and by rotating the gaugino and Higgs fields, one
can eliminate the phases in all but $\mu$ and the $A$'s \cite{dgh}.

\begin{figure}[htb]
\epsffile{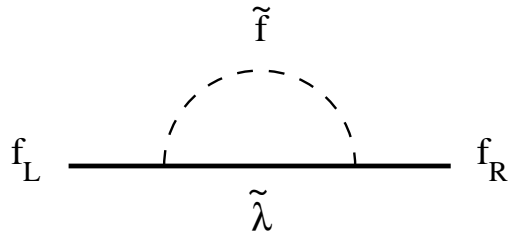}
\vspace{-1.0cm}
\caption{Diagrams contributing to quark and electron EDM's}
\label{feyndiag}
\end{figure}

The electric dipole moments of the quarks and electron receive SUSY
contributions from the diagrams of Figure \ref{feyndiag}.  Here
$\tilde\lambda$ can be either a gluino $\tilde g$, chargino
$\widetilde W$ or neutralino $\tilde\chi^0$, and it is understood that
an external photon line attaches to either the internal sfermion or
$\char$ line.  The necessary $\cp$-violation either accompanies the
mixing between left and right-handed sfermions or arises from the
mass/mixing matrices for the $\char$'s or $\neut$'s (due to the
presence of $\mu$ in both mass matrices).

Full expressions for the $\char$, $\neut$ and $\gluino$ exchange
contributions to the quark and electron EDM's in terms of the SUSY
parameters can be found in \cite{ko}.  The $\gluino$ exchange
contribution to the quark EDM takes a particularly simple form:
\begin{equation}
  \label{dg}
  d_q^g/e\sim {\alpha_s\over\pi}{m_q m_{\tilde g}
    |A^*+\mu\tan\beta|\over\msf^4} \sin\gamma
\end{equation}
For up-type quarks, take $\tan\beta\to\cot\beta$.  Here $\gamma$ is
the argument of the off-diagonal element of the squark mass matrix,
$\gamma=\arg(A^*+\mu\tan\beta)$.  For typical values of the masses,
$m_{\tilde g}=\msf=|A^*+\mu\tan\beta|=100\gev$, the requirement that
the quark EDM contribution to the neutron EDM satisfy the experimental
bound\cite{nexp} of $|d_n|<1.1\times10^{-25}e\:{\rm cm}$ implies that the phase
$\gamma$ be very small, $\sin\gamma\la 0.001$.  However, this bound
can be considerably relaxed by making the squarks heavier.  The
$\char$ exchange contribution also has a simple dependence on the
SUSY phases, as $d_q^c/e\sim\sin\thm$, while the $\neut$ exchange
contribution has a more complicated dependence.  Finally, we use the
non-relativistic quark model to relate the neutron EDM to the up and
down quark EDM's via $d_n=(4d_d-d_u)/3$.

We recall that a general neutralino is a linear combination of the
neutral guaginos and higgsinos, $\tilde\chi^0_i=\alpha_i\widetilde W_3
+ \beta_i\bino + \gamma_i \widetilde H_1+\delta_i \widetilde
H_2$. For large $\mu > M_2, M_Z$, however, the lightest neutralino is
very pure bino. We consider the case of a bino as the lightest
supersymmetric particle (LSP).  To compute the $\widetilde B$ relic
density, we calculate the $\bino$ annihilation cross-section;
$\Omega_{\bino} h^2\sim\langle\sigma_{\rm ann}v_{\rm
  rel}\rangle^{-1}$.  $\bino$ annihilation is dominated by
sfermion exchange into fermion pairs.  This process exhibits ``p-wave
suppression''; that is, the zero-temperature annihilation rate is
suppressed by powers of the final state fermion mass.  Note that
raising $\msf$ turns off this annihilation channel, and so bounding 
$\Omega_{\bino} h^2$ places an upper limit on the sfermion masses as
well as on the $\bino$ mass.

It has been shown\cite{fkos} that $\Omega_{\bino} h^2$ may in some
cases be sensitive to the presence of $\cp$-violating phases in the
sfermion mass matrix.  Since
$\bino$'s freeze out when they are non-relativistic, it is convenient
to expand the annihilation cross-section $\langle\sigma_{\rm ann}v_{\rm
  rel}\rangle = a + b(T/\mb)+\ldots$.  In the absence of
$\cp$-violation and sfermion mixing, and taking ${\msf}_1={\msf}_2$,
$a$ is given by 
\begin{equation}
  \label{anocpv}
  a_f = {{g'}^4\over
    128\pi}\,(Y_L^2+Y_R^2)\,{m_f^2\over({\msf}^2+{\mb}^2-m_f^2)^2},
\end{equation}
and the p-wave suppression is evident, as $a_f\sim m_f^2$. Here
$Y_L$($Y_R$) is the left(right) sfermion hypercharge.  In the
presence of $\cp$-violation and sfermion mixing, and taking
${\msf}_1\approx {\msf}_2$,
\begin{eqnarray}
  \label{ayescpv}
  a_f &=& {{g'}^4\over 32\pi}\,Y_L^2 Y_R^2\,{{\mb}^2\over({\msf}^2+
      {\mb}^2-m_f^2)^2}\times\nonumber\\
      & & \;\;\;\sin^22\theta_f\sin^2\gamma_f\;+\;O(m_f \mb),  
\end{eqnarray}
where $\theta_f$ is the mixing angle between left and right sfermions,
and $\gamma_f$ is the phase described above.  In this case, $a_f$
contains a piece which is {\em not} p-wave suppressed.

In this talk, I will consider the case of the Constrained MSSM
(CMSSM).  In this \"Ansatz, the scalar masses are taken equal to a
universal $m_0$ at a unification scale $M_X$, the gaugino masses unify
to $\m12$, and the trilinear couplings $A_f$ are set equal to
$A_0$.  The renormalization group equations are then used to run the
parameters down to the electroweak scale.  We are left with two
independent phases, $\thm$ and $\tha$ at the scale $M_X$.

\begin{figure}[htb]
\epsffile{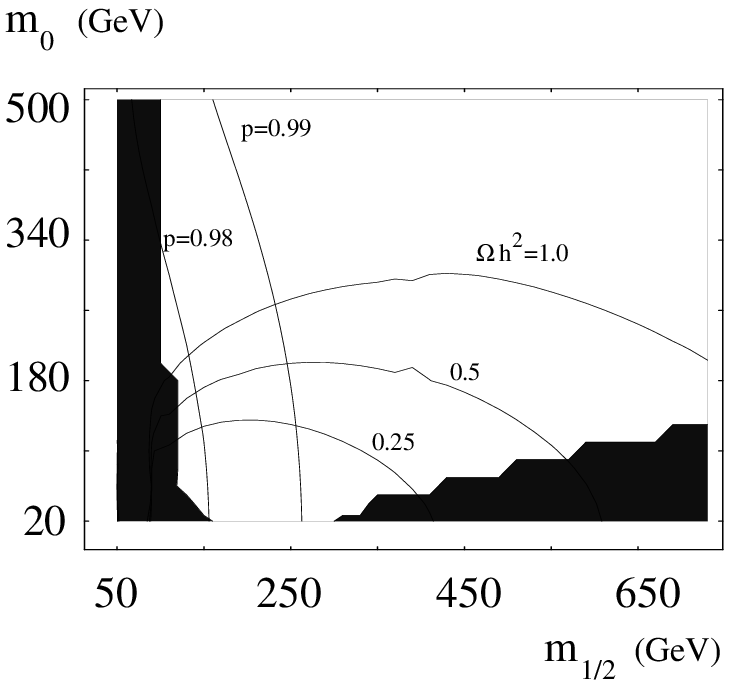}
\vspace{-1.5cm}
\caption{}
\label{ohsq}
\end{figure}

In Figure \ref{ohsq} we display contours of constant
$\ohsq=0.25, 0.5,$ and $1.0$, as a function of $m_0$ and $\m12$, for
$\tan\beta$=2.1, and taking $A_0=300\gev, \tha=0.8\pi$ and $\thm=0$
\cite{fko}.  The dark regions are excluded because they produce either
light $\char$'s or light sfermions or lead to staus or stops as the
LSP.  Also plotted are curves of constant $\bino$ purity, and we
observe that the neutralinos near their mass upper bound for
$\ohsq=0.25$ are very pure ($p>0.99$) bino, so that the $\neut$'s
will annihilation predominantly through sfermion exchange, as
described above.  Requiring $\ohsq\le 0.25$, the resulting upper bound
on $\m12$ is $\approx 400\gev$, corresponding to $m_{\bino}\la 160\gev$.

This bound is quite independent of the parameters $A_0, \tha$ and
$\thm$.  Recall that the lifting of the p-wave suppression described
above requires both $\cp$-violation and significant sfermion mixing
(though it can be lifted to some extent by sfermion mixing alone).
Since annihilation into leptons is particularly enhanced, we consider
stau mixing.  At the electroweak scale, the left and right stau mass
parameters are split by $m_L^2-m_R^2\approx0.4\m12^2$. Then for the
part of the zero temperature cross-section which is not p-wave
suppressed, $\langle\sigma_{\rm ann}v_{\rm
  rel}\rangle_{T=0}\sim\sin^22\theta_\tau$, where
\begin{equation}
\sin^22\theta_\tau\approx
0.01\left({100\gev\over\m12}\right)^4\left({A^*+\mu\tan\beta\over
100\gev}\right)^2
\end{equation}
This is very small for $\m12$ near its upper bound, and so the zero
temperature annihilation rate remains suppressed.

The neutron EDM is sensitive to the masses of the squarks, which in
the CMSSM are given by $m_{\tilde q}^2\approx
m_0^2+6{\m12}^2+O(m_Z^2)$.  These masses, and consequently the neutron
EDM, are insensitive to $m_0$ in the cosmologically allowed region
(see Figure \ref{ohsq}).  We also find that the dominant contribution
to the neutron EDM comes from $\tilde W$-exchange (unless
$\tha\gg\thm$), and so is insensitive to $A_0$.  Minimizing the
neutron EDM is then achieved by taking $\m12$ as large as is
cosmologically allowed.

\begin{figure}[htb]
  \vspace{-1.3cm}
  \epsffile{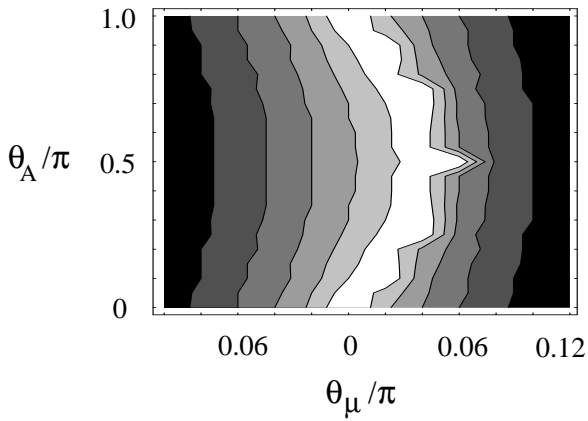}
  \vspace{-1.0cm}
  \caption{Minimum value of $\m12$ needed to bring the neutron EDM
    below experimental bounds.}
  \label{neutedm}
\end{figure}

\vspace{-0.5cm} Accordingly, in Figure \ref{neutedm} we plot, as a
function of $\thm$ and $\tha$, the {\em minimum} value of $\m12$
needed to bring the neutron EDM down below its experimental bound of
$1.1\times 10^{-25}\,e\:{\rm cm}$.  The light central region has
$\m12^{\rm min}<200\gev$, and successive contours represent steps of
$100\gev$.  The black  regions yield a stau as the LSP.  Since $\char$
exchange dominates unless $\tha\gg\thm$, and since $\tha$ contributes
only part of $\gamma$, the neutron EDM is fairly insensitive to
$\tha$.  The contours are bowed to the right of $\thm=0$, where there
is a cancellation between the $\char$ and $\gluino$ exchange
contributions.  There are also similar allowed regions near $\thm=\pi$
and for negative $\tha$.  Recalling that $\ohsq<0.25$ requires
$\m12\la400\gev$, we see from Figure \ref{neutedm} that
$|\thm|\la\pi/10$, while $\tha$ is essentially unconstrained.

\begin{figure}[htb]
  \vspace{-1.0cm}
  \epsffile{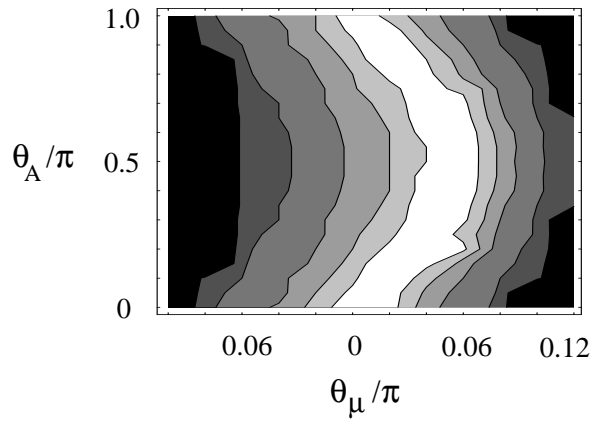}
  \vspace{-1.0cm}
  \caption{Same as Figure \ref{neutedm} for the e$^-$ EDM}
  \vspace{0.2cm}
  \label{elecedm}
\end{figure}

The above bounds may be sensitive to the spin
structure of the nucleon\cite{ef}, so it is important to also consider
bounds from the electron EDM.  In Figure \ref{elecedm}, we require
$\m12$ to be large enough so that the e$^-$ EDM is less than
$1.9\times 10^{-26} e\:{\rm cm}$\cite{eexp}.  We find the bounds on
$\thm$ from the e$^-$ EDM are comparable to those from the neutron
EDM.

I would like to gratefully acknowledge Keith Olive and Mark Srednicki,
with whom this work was done in collaboration.

\end{document}